\documentclass[sn-basic]{sn-jnl}

\usepackage{geometry}
\usepackage{comment}
\usepackage{graphicx}%
\usepackage{multirow}%
\usepackage{amsmath,amssymb,amsfonts}%
\usepackage{amsthm}%
\usepackage{mathrsfs}%
\usepackage[title]{appendix}%
\usepackage{xcolor}%
\usepackage{textcomp}%
\usepackage{manyfoot}%
\usepackage{booktabs}%
\usepackage{algorithm}%
\usepackage{algorithmicx}%
\usepackage{algpseudocode}%
\usepackage{listings}%
\usepackage{color}
\usepackage{siunitx}

\begin{document}

\title[Article Title]{Detecting Phishing in Ethereum Networks using Quantum Machine Learning}

\author*[1]{\fnm{Sai Sakunthala} \sur{Guddanti}}\email{saisakunthala1204@gmail.com}

\author[2]{\fnm{Anupama} \sur{Ray}}\email{anupamar@in.ibm.com}

\author[3]{\fnm{Mrunal}
\sur{Arun Kumavat}}\email{mrunalkumavat23@gmail.com}

\author[4]{\fnm{Anil} \sur{Prabhakar}}\email{anilpr@ee.iitm.ac.in}

\affil[1]{\orgdiv{Centre for Quantum Info. Comm. and Computing}, \orgname{Indian Institute of Technology Madras}, \city{Chennai}, \country{India}}
\affil[2]{\orgdiv{IBM Research}, \orgname{IBM},  \city{Bengaluru},  \country{India}}
\affil[3]{\orgdiv{Dept. of Physical Sciences}, \orgname{Indian Institute of Science Education Research, Mohali},  \country{India}}
\affil[4]{\orgdiv{Electrical Engineering}, \orgname{Indian Institute of Technology Madras},  \city{Chennai},  \country{India}}


\abstract{This article explores the potential of Quantum Machine Learning (QML), specifically assessing a Quantum Support Vector Machine (QSVM) and a Variational Quantum Classifier (VQC) for detecting anomalies in real-world financial transaction data. While these QML methods outperform statistical methods, they fall short of cutting-edge deep learning techniques. To bridge this gap, we propose a hybrid quantum-classical ensemble framework that leverages the strengths of both domains. We demonstrate its effectiveness in detecting phishing in Ethereum transaction networks by combining complementary algorithms. The QSVM, whether used individually or in an ensemble, consistently delivered the lowest false negatives and higher recall rates, that are crucial for anomaly detection. To enhance individual models, we encoded the data using novel cascaded Quantum Random Access Coding (QRAC) schemes and compared it with the popular encoding ZZ feature map on both simulators and the IBM Heron quantum processor. For both QSVM and VQC, we consistently observed improvements (13\% for QRAC-VQC and 3\% for QRAC-QSVM) of QRAC over the ZZ feature map. Notably, certain QML algorithms exhibit remarkable resilience on the IBM Heron quantum processor, approaching simulator-level performance on devices with high quantum volume. This observation underscores the promise of QML despite hardware limitations.}




\maketitle

\section{Introduction}\label{sec1}

Quantum Machine Learning (QML) is one of the most promising directions where quantum computing principles are expected to make an impact in the near future. This potential arises from the interplay between machine learning and quantum computing, where ideas from one domain can enhance the other. For instance, quantum kernels can improve ML algorithms, while classical ML techniques can aid error mitigation and quantum hardware optimization. While theoretical results indicate that QML algorithms can outperform classical ones for certain problems \citep{Liu_2021}, practical demonstrations of quantum advantage remain limited due to constraints in hardware, data encoding, training, and other challenges.

In this work, we benchmark two QML algorithms, namely, Variational Quantum Classifiers (VQC) and Quantum Support Vector Machines (QSVM). For VQCs, we evaluate performance using the ZZ feature map and QRAC. For QSVMs, we compare a gate-based implementation (QSVM\_qiskit) with a QRAC-based encoding method, highlighting the advantages of QRAC for binary classification tasks with annealing based implementation of QSVM.

We compare QML performance against state-of-the-art classical machine learning (ML) models, and we observe complementary strengths between QSVM and classical algorithms for specific tasks, such as phishing detection. Motivated by this, we propose a hybrid quantum-classical ML ensemble that leverages the strengths of both paradigms to enhance classification accuracy.

We analyze the learning ability of algorithms on IBM statevector simulators and real quantum processing units (QPUs). This allows us to validate our methods on noisy intermediate-scale quantum (NISQ) hardware and bridge the gap between simulations and real-world applications.  Our contributions are
\begin{itemize}
\item design of a hybrid quantum-classical ensemble framework for phishing detection using the stacking technique,
\item benchmarking QML algorithms (VQC and QSVM) with QRAC-based encoding,
\item performance evaluation on IBM quantum hardware and statevector simulators to assess scalability and feasibility,
\item novel cascaded QRAC encoding scheme using (2,1)-QRACs,
\item application of QML techniques to phishing detection
\end{itemize}

The methods and results presented in this work are versatile and applicable to a wide range of classification tasks. By systematically combining simulations and hardware experiments, we ensure the reliability of our findings and provide a robust foundation for future advancements in QML. In an earlier version of this work, we had discussed the proposed ensemble framework~\citep{ray2022classical}. In this revised manuscript, we present a more focused analysis.

\section{Related work}
There are several machine learning tasks where quantum computers hold the promise of improvements over classical systems \citep{BLSF19, BCK+22}. Classification problems in QML are primarily approached through variational models \citep{MNKF18, HCT+19, SBSW20} or kernel-based methods \citep{HCT+19, SK19}. While quantum methods have been explored for regression tasks \citep{MNKF18, PEK22} and solving equations \citep{XSE+21, LJM+20}, their practical utility remains under development due to hardware and scalability limitations.

Anomaly detection using quantum algorithms is a nascent area. In credit card fraud detection, \cite{ZLW21} proposed a variational quantum Boltzmann machine, while \cite{HOR21} developed a variational quantum-classical Wasserstein GAN. Recent work by \cite{Hdaib2024} explores quantum deep learning (QDL), emphasizing the use of hybrid quantum-classical models. \cite{Innan2024} demonstrated that Quantum Graph Neural Networks (QGNNs) outperform classical models in financial fraud detection.

For phishing detection, classical research has made significant strides \citep{MWX+21, PSCH21,KIJ13}. Specifically, \cite{EtherACMToIT} achieved state-of-the-art results using a graph convolutional network (GCN) trained on transaction graph features from the Ethereum network. Their method leverages eight financial features to identify phishing nodes effectively \citep{CGCZL20}.

Literature also shows that ensembles of algorithms often outperform individual models, and recent works in QML propose quantum ensembles. For instance, \cite{Maria} introduced a quantum equivalent of Bayesian model averaging, and \cite{Macaluso2} extended bagging concepts to quantum ensembles. However, practical implementation of such methods is challenging on noisy intermediate-scale quantum (NISQ) devices. In this paper, we propose a stacking-based hybrid quantum-classical ensemble framework that leverages the complementary strengths of classical ML and QML algorithms.

One key aspect of our work is the use of Quantum Random Access Coding (QRAC) for encoding classical data into quantum states. QRAC has shown promise in reducing the parameter count in VQC models \citep{9511252} and improving data embedding efficiency. \cite{QRAC2009} highlighted the foundational advancements in QRAC, exploring its applications in communication complexity and locally decodable codes. Building on these insights, we apply QRAC for binary classification tasks, demonstrating its potential to enhance phishing detection in Ethereum networks.

\section{QML algorithms: {Background}}
In this section, we describe two quantum algorithms - QSVM and VQC that have been used in this article. For any quantum model to work on classical data, the first step is to be able to load the classical data as quantum states. This step is called \textit{data encoding or mapping}, and we use parameterized quantum circuits as feature maps. We have experimented with the ZZ feature map and the QRAC. To load classical features as quantum states we can apply different types of transformations on qubits, such that the qubits of a quantum circuit now represents each data point and its features.

\subsection{Quantum Random Access Coding (QRAC)}
QRACs can be applied to reduce the number of qubits required to represent binary or categorical features. They allow us to encode $n$ bits into $m$ qubits, for $n>m$, so that any one of the bits can be extracted with a success probability at least $p>1/2$. The $(n,m,p)$-QRAC is a feature map that maps a $n$-bit classical string to $m$ qubits, such that the probability of recovering any one classical bit is at least $p$. The efficient QRACs for encoding up to 3 bits into one qubit are known: (2,1,0.85)-QRAC for encoding 2 classical bits into 1 qubit and (3,1,0.78)-QRAC for encoding 3 classical bits into 1 qubit. For ease of representation, we write them as (2,1)-QRAC and (3,1)-QRAC. All necessary unitary gates and measurements can be realized by the $U_3$ rotation gates. 

\begin{figure}[htb]
\centering
\includegraphics[width = 0.6\linewidth]{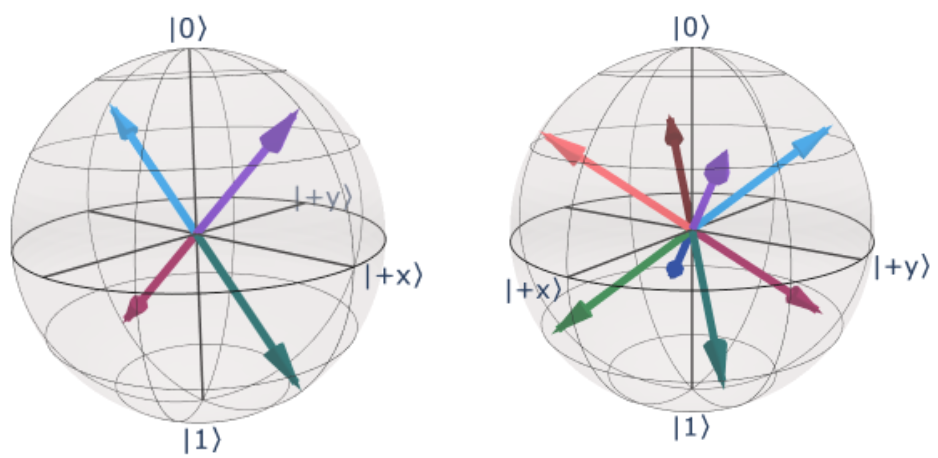}
\caption{The four states of (2,1)-QRAC as corners of square (left) and the eight states of (3,1)-QRAC as corners of cube (right).}
\label{qracs}
\end{figure}

For (2,1)-QRAC encoding, all the four quantum states representing the four 2-bit classical strings ('00' ,'01', '10', '11') are such that they are at a maximum distance from each other. In Figure \ref{qracs}, we observe that all four states lie on the edges of square. Similarly for (3,1)-QRAC encoding, all the eight quantum states representing the 3-bit classical strings ('000', '001', '010', '011', '100', '101', '110', '111') lie on the corners of a cube on the Bloch sphere, as it maximizes the distances between them, also shown in Figure \ref{qracs}. \cite{4036000} showed that the (4,1)-QRAC is not possible, hence we consider up to (3,1)-QRACs only

For our work, we focus on binary feature embeddings using QRACs. Real-valued features are thresholded to convert them into binary values, where exceeding the threshold maps the feature to '1', otherwise to '0'. While categorical embeddings can also be achieved by grouping real-valued features into distinct categories, binary embedding was chosen for simplicity and effective utilization of QRAC capabilities. A more detailed exploration of QRAC-based embeddings, including categorical feature embeddings, is available in our related work \citep{10313693}.

To optimize qubit usage, concatenated QRACs are employed. For instance, the third 3-bit string QRAC can be added to the first or second qubit, reducing the encoding requirement from three qubits to two. This approach enables all seven features to be encoded using just two or three qubits.

However, this concatenation approach may introduce certain issues during training. Although the concatenated version represents two QRACs using a single qubit, it can sometimes cause one QRAC to dominate, effectively nullifying the contribution of the other due to excessive overlap between their encoded states. To overcome this limitation, we propose a novel cascaded QRAC architecture (see Section~\ref{new cascaded}), building upon our previous work \citep{10313693}.

\subsubsection{New cascaded QRAC encoding} \label{new cascaded}

\begin{figure}[h]
    \centering
    \includegraphics[width=0.65\linewidth]{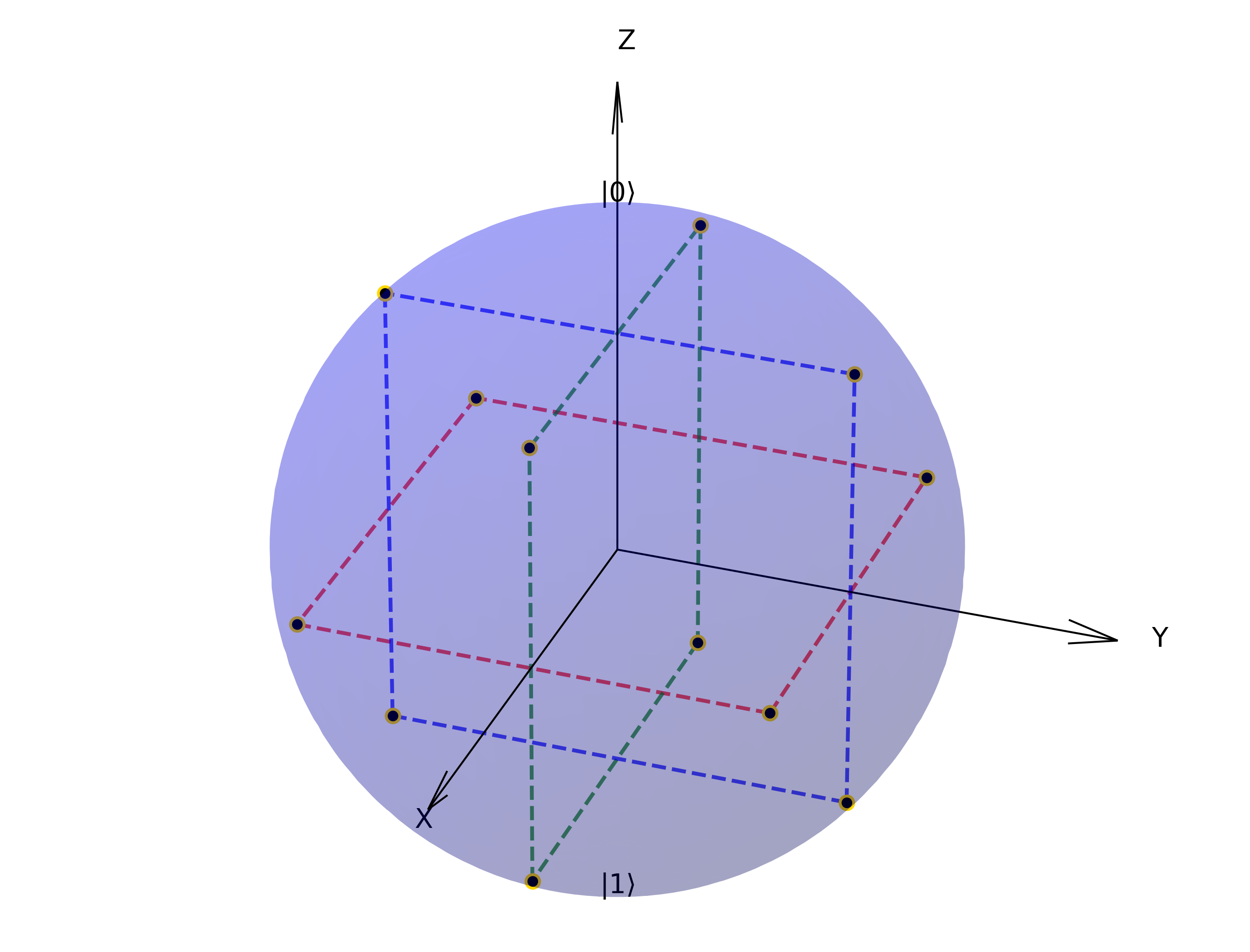}
    \caption{The feature encoding scheme of 2 qubit (2,1)-QRAC, based upon mutually orthogonal plane rotations}
    \label{fig:bloch}
\end{figure}

To improve the separation between features, we use (2,1)-QRAC style encodings on three orthogonal planes of the Bloch sphere: the $xy$, $yz$, and $xz$ planes, as shown in Figure \ref{fig:bloch}. Each feature is assigned to one of these planes, and rotations are chosen based on a small fixed set of points on each plane. This does not guarantee full independence between features, since some combinations of rotations can lead to overlapping points. However, separating features across orthogonal planes reduces interference compared to concatenating two QRACs on a single qubit. This method increases the effective independence between features and helps preserve their distinguishability in the quantum state.

Once we have loaded the data, we follow different flows for QSVM and VQC.

\subsection{Variational Quantum Classifier (VQC)}
The VQC method involves encoding classical data into quantum states, followed by parameterized quantum circuits (also called ansatz) to learn the mapping. The measurement results are post-processed using parity checks, with even parity labeled as +1 and odd parity labeled as -1. The classical optimizer adjusts the parameters of the ansatz to minimize the cost function, similar to training a neural network. We used COBYLA, a gradient-free optimizer, for this task, experimenting on IBM's noiseless state vector simulator before conducting experiments on quantum processing units (QPUs).

\subsection{QSVM using annealing (QSVM\_dwave)}
Quantum annealing is a heuristic approach to adiabatic quantum computing that leverages quantum tunneling to escape local optima, unlike classical annealing, which relies on thermal fluctuations. Quantum annealers, consisting of superconducting qubits and couplers, solve problems expressed in QUBO form. \cite{2} formulated QUBOs for different ML models. In this work, we formulate the classical SVM optimization problem as a QUBO. Given training data $X \in R^{N \times d}$ and training labels $Y \in \{-1,+1\}^N$, the hyperplane is determined by weights, $w\in R^d$, and bias, $b\in R$, that separate the training data into binary classes. Mathematically,
\begin{equation}
        \min_{w,b} \;\frac{1}{2}\|w\|^2, \quad
    \text{subject to} \quad y_i(w^Tx_i+b) \geq 1,\;
    \forall \;i\in\{1,2,\dots,N\}. \nonumber
\end{equation}
\\Converting this to dual form and incorporating a kernel yields the QUBO. 
\begin{align}\label{eqn:dual}
    \min_{\lambda}\mathcal{L}(\lambda)=\frac{1}{2}\lambda^T(K\odot YY^T)\lambda-\lambda^T 1_N, \; \lambda\geq 0_N,
\end{align}

\noindent Where the kernel matrix for the Radial Basis Function (RBF) kernel is defined as $K_{ij}=e^{-\lVert x_i-x_j \rVert^2/2\sigma^2}$ (with $\sigma=150$, to have the kernel elements $\in$ [0,1]) and $\odot$ is the element-wise multiplication operation. 

The QUBO matrix is provided as input to a quantum annealer, which solves the minimization objectives and returns the Lagrange multipliers (binary), which are the support vectors. To extend the support vectors beyond binary values, we introduce a precision vector $p = [2^0,2^1]$, doubling the QUBO dimensions and allowing integer support vectors in the range [0,3]. The modified QUBO is minimized using the annealer, returning the optimal $\hat{\lambda}$ from which the final support vector values $\lambda$ are computed

The prediction for unseen data follows:
\begin{align}\label{eqn:pre}
    \mathsf{label}(x)  &= \mathrm{sign} \left(\sum^{N}_{i=1} \lambda_iy_i(K_{xi})+b\right),\\
     b  = \mathsf{mean}&(y_i-w^Tx_i),\;\;\;
     \text{where}\;\;w^Tx_i=\sum^{N}_{j=1}\lambda_jy_jK_{ji}
\end{align}

\subsection{QSVM using quantum circuits (QSVM\_qiskit)}
A quantum kernel is the inner product of quantum feature maps. We expect to see a quantum advantage when we choose a quantum feature map that is not easy to simulate on a classical computer \citep{HCT+19}. We first initialize a circuit in the $\vert 0\rangle$ state, then we apply feature maps $\phi(\Vec{x_i})$ and $\phi^\dagger(\Vec{x_j})$ to the circuit, which changes the state to $\phi^\dagger(\Vec{x_j})\phi(\Vec{x_i})\vert 0\rangle$. We then measure the circuit in Z bases and obtain the kernel element as the probability of measuring $\vert 0\rangle$. Mathematically, kernel elements can be represented as
\begin{align}
    K(\vec{x_i},\vec{x_j}) =K_{ij}=& \mathrm{Pr}[\mathsf{measure}\;\vert 0\rangle]
    = \vert\langle 0\vert \phi^\dagger(\vec{x_j})\phi(\vec{x_i})\vert 0\rangle\vert^2\nonumber
    = \vert\langle\phi^\dagger(\vec{x_j})\vert\phi(\vec{x_i})\rangle\vert^2 .\nonumber 
\end{align}

\noindent We finally use the quantum kernel matrix to train the SVM classically. 
\begin{enumerate}
    \item substitute the kernel matrix in the dual form of SVM, as in (\ref{eqn:dual})
    \item Perform an optimization same as we would for a classical SVM, and predict the data using (\ref{eqn:pre}).
\end{enumerate}

\section{Dataset and feature extraction}
\begin{table}[htb]
    \centering
    \caption{Average and standard deviation of the feature values of phishing (1160 nodes) and non-phishing (10800 nodes) nodes of the dataset used.}
    \label{tab:mean_and_sd}

    \begin{tabular}{|l|S|S|S|S|S|S|S|}
    \hline
    {Statistical measure} & {In-degree} & {Out-degree} & {Degree} &
    {In-strength} & {Out-strength} & {Strength} & {Neighbors} \\
    \hline
    Phishing avg          & 31.39  & 20.49  & 51.88  & 78.61   & 86.73   & 165.34  & 31.69 \\
    Non-phishing avg      & 4.50   & 4.64   & 9.14   & 72.53   & 9.25    & 81.78   & 3.64  \\
    Phishing ($\sigma$)   & 180.99 & 96.83  & 219.73 & 691.29  & 860.10  & 1390.35 & 106.39 \\
    Non-phishing ($\sigma$) & 154.35 & 101.32 & 192.20 & 4409.68 & 281.02 & 4421.54 & 79.36 \\
    \hline
    \end{tabular}
\end{table}

\label{sec:dataset}

The Ethereum blockchain platform has grown rapidly, and with it, phishing activities targeting blockchain security. Research on anomaly detection in blockchain networks has been limited, mainly due to a lack of labeled data, space-time constraints, and severe class imbalance. We crawled 3 million Ethereum nodes, with 1160 phishing nodes (0.039\%) and 2,972,324 non-phishing nodes. Phishing nodes are identified from public reports on Etherscan\footnote{https://etherscan.io}. The dataset is highly imbalanced, with non-phishing nodes outnumbering phishing nodes by 2500 times, posing a challenge for prediction algorithms.

We use 7 statistical features from \cite{EtherACMToIT}, which include in-degree, out-degree, degree, in-strength, out-strength, strength, and the number of neighbors. Phishing nodes show significant discrepancies in these features: they have higher transaction values (strength), more neighbors, and extreme values in degree and strength. These features are indicative of phishing behavior, as phishing nodes often have more transactions and interactions compared to non-phishing nodes. Table \ref{tab:mean_and_sd} shows the mean and standard deviation for phishing and non-phishing transactions. While the mean feature values differ considerably, the high standard deviation reflects the challenge of classification, indicating the variability in node behaviors within each class.

\section{Methodology}
We have investigated classical and quantum models for benchmarking phishing detection and explored a hybrid quantum-classical ensemble. To ensure a fair comparison with quantum models, all classical models also used 320 training samples (160 phishing and 160 non-phishing) and 11,000 testing samples (1000 phishing and 10,000 non-phishing). The training set was restricted to 320 samples to construct a balanced dataset. Since only 1160 phishing nodes were available, 1000 were reserved for testing and the remaining 160 were used for training. An equal number of non-phishing samples were then selected, resulting in a balanced training set containing 160 phishing and 160 non-phishing samples.

Classical methods included Graph Convolutional Networks (GCN), Light Gradient Boosting Machine (LGBM), and SVM. GCNs, the state-of-the-art for graph problems, demonstrated superior performance but with significant computational demands. Quantum algorithms benchmarked included Variational Quantum Classifiers (VQC) and Quantum Support Vector Machines (QSVM) with both quantum kernels and annealing-based formulations.
\subsection{Ensembles of hybrid Quantum-Classical ML Models}
We implemented an ensemble technique to combine the strengths of quantum and classical models. Bagging showed no improvement and is hence excluded from our discussion.
\subsubsection{Stacking:}
\begin{figure}[htb]
    \centering
    \includegraphics[width = 0.9\linewidth]{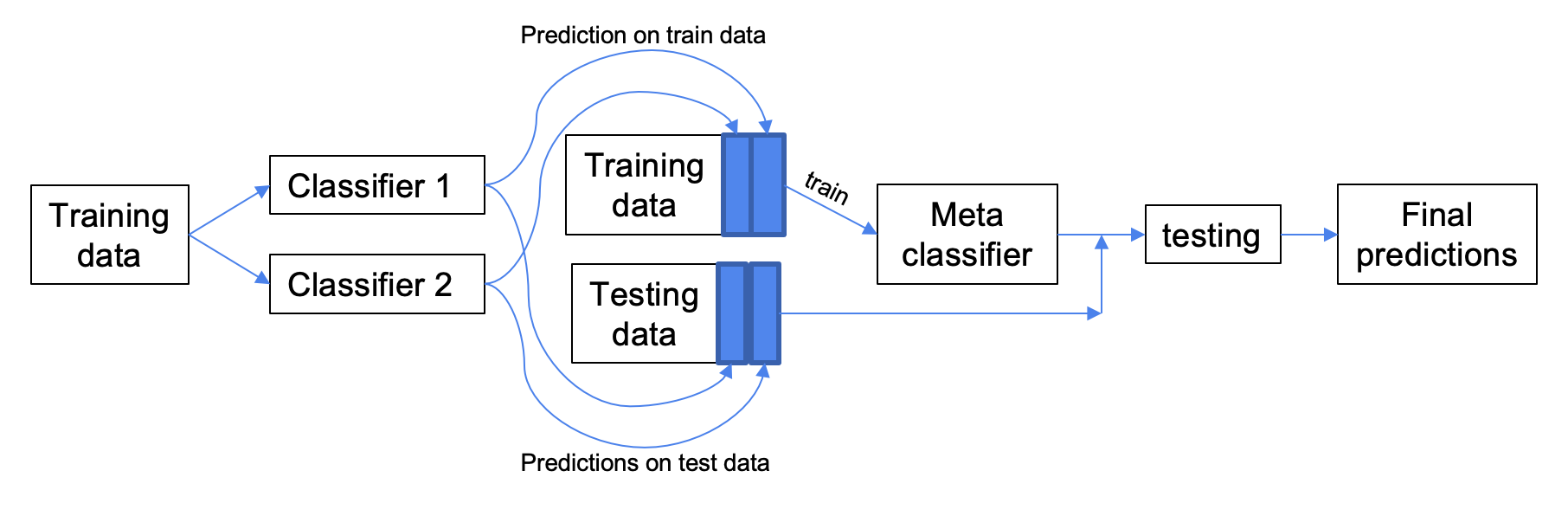}%
    \caption{Two level stacking technique implemented}%
    \label{stacking}%
\end{figure}
The process of two-level stacking is illustrated in Figure \ref{stacking}. In this method, base models generate prediction vectors from the training data, which are then appended as additional features to the original training set. These enriched features are used to train a meta-classifier. For consistency, the base models also generate prediction vectors for the testing data, which are appended to the testing set and used by the meta-classifier for final predictions.

To combine complementary algorithms we chose the model with the least false negatives (QSVM\_qiskit) and the overall best performing model (GCN).

\subsection{Framework for QRAC usage}
As previously detailed in \citep{10313693}, to use QRACs, we convert real-valued features into binary data, with each feature binarized by comparing its value to the median across the dataset, assigning values as either '0' or '1.' The binarized features are then grouped into 3-bit strings: in-degree, out-degree, and degree (all degree features) form the first string; in-strength, out-strength, and total strength (all strength features) form the second; and the number of neighbors is padded to create the third string. These 3-bit strings are encoded using three (3,1)-QRACs and mapped to three qubits.

Using the novel cascaded QRAC method, we represent seven binarized features as four 2-bit strings (with the final one padded with a zero), yielding four total encodings. These encodings are distributed as follows: four features are mapped onto one qubit using two orthogonal planes, while the remaining three features are mapped onto a second qubit, also using two orthogonal planes. This arrangement prevents feature overcrowding and maintains the distinctiveness of each encoding without introducing additional qubits. We refer to this scheme as the 2-qubit (2,1)-QRAC encoding, or 2q (2,1)-QRAC for short.

Once encoded, the QRAC-based features are used in different quantum algorithms. This framework provides an efficient method for QRAC usage while minimizing qubit requirements.

\subsection{Platform for Hardware Experiments}

The quantum machine learning models, QSVM\_qiskit and VQC, were trained using Qiskit Aer simulators and evaluated on IBM Quantum hardware using the Qiskit Runtime primitives framework \cite{qiskit_runtime}. Training was performed entirely on simulators to enable efficient iterative optimization and parameter estimation while avoiding the substantial queue latency and execution overhead associated with training directly on quantum hardware.

Hardware inference for the VQC model was performed using the \texttt{SamplerV2} Runtime primitive. The trained feature map and ansatz circuits were composed, transpiled for the target backend, and parameterized with both input features and trained variational weights before execution on hardware. Multiple parameter-bound circuits corresponding to batches of test samples were grouped and submitted together within a single Runtime job, improving hardware utilization and reducing communication and queue overhead. Classification predictions were obtained from measured bitstring outcomes using a parity-based decision rule derived from the measured shot distributions.

For QSVM\_qiskit, hardware inference was performed using fidelity-based quantum kernel evaluations implemented through \texttt{ComputeUncompute} and \texttt{FidelityQuantumKernel}. Since QSVM inference depends only on the support vectors identified during simulator-based training, hardware kernel evaluations were restricted to these support vectors, significantly reducing the number of required quantum circuit executions. Kernel evaluations between batches of test samples and the support vectors were executed on hardware using Runtime primitives, and the resulting kernel matrices were used to compute the QSVM decision function classically.

All hardware-executed quantum circuits were transpiled using Qiskit preset pass managers to generate ISA-compliant circuits compatible with the selected IBM Quantum backend. Transpilation also optimized circuit execution through backend-aware gate decomposition and circuit optimization passes, reducing execution depth and improving hardware compatibility.

\section{Results and Analysis}

This section presents the results using individual and ensemble models. We trained QSVM\_qiskit and VQC on IBM statevector simulator using Qiskit and we trained QSVM\_dwave on Dwave hybrid solver from Leap cloud.
\subsection{\large{Best individual model results}}
\begin{figure}[htb]
    \centering
    \includegraphics[width = 0.75\linewidth]{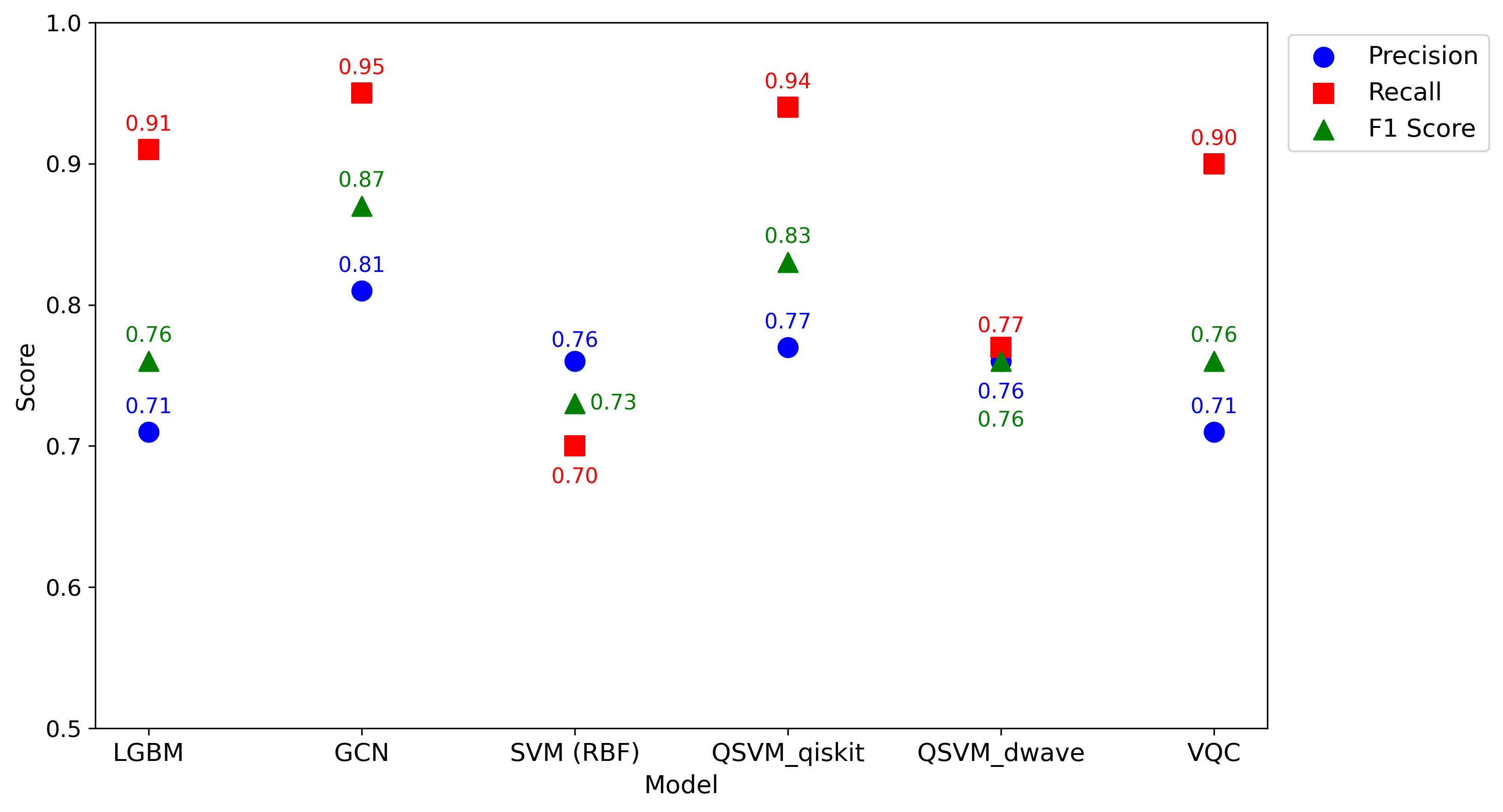}%
    \caption{Macro average of Precision, Recall and F1 scores of individual classical and quantum models trained using same train (160p and 160np) and test data (1000p and 10000np).}%
    \label{fig:indmodel}%
\end{figure}
Figure \ref{fig:indmodel} shows the macro average of precision, recall and F1 score of each individual quantum and classical algorithms. The results of models VQC and QSVM\_qiskit that are in the graph, are obtained by using (3,1)-QRAC as feature map, for VQC using 2 qubit (3,1)-QRAC and for QSVM using 3 qubit (3,1)-QRAC, because they gave best results compared with other feature embeddings.

Graph Convolutional Networks (GCN), the state-of-the-art for graph problems, achieved high performance but required $\sim$17 hours ($\sim$8K epochs) to train on the full graph. Light Gradient Boosting Machine (LGBM) and Support Vector Machines (SVM) used extracted features from the graph. LGBM outperformed other classical models in recall, although with low precision, and required under 2 minutes ($< 10$ epochs) for training.

The performance of QSVM\_qiskit is only slightly less when compared to state-of-art classical GCN. GCN can take the entire graph during its training, while all other models are limited to the 7 classical features extracted. To ensure a fair comparison with the quantum models, we trained a classical SVM and a LGBM using the same 7 features. We consistently observed better results with QSVM\_qiskit using QRACs compared to both these classical models. Also VQC using QRAC is at par with LGBM. These results suggest that QRAC-based encodings can improve the practical performance of current QML models.

\subsection{\large{Ensemble results}}
We observe that for the phishing class, QSVM\_qiskit has the lowest number of false negatives and highest number of true positives, which is best. Thus when used in an ensemble, QSVM would contribute towards increasing phishing recall. This is very important for the application in consideration as we are still okay with some non-phishing transactions being flagged as phishing, while we need to reduce actual phishing transaction getting mislabeled as non-phishing (false negatives)

\begin{figure}[htb]
    \centering
    \includegraphics[width = 0.6\linewidth]{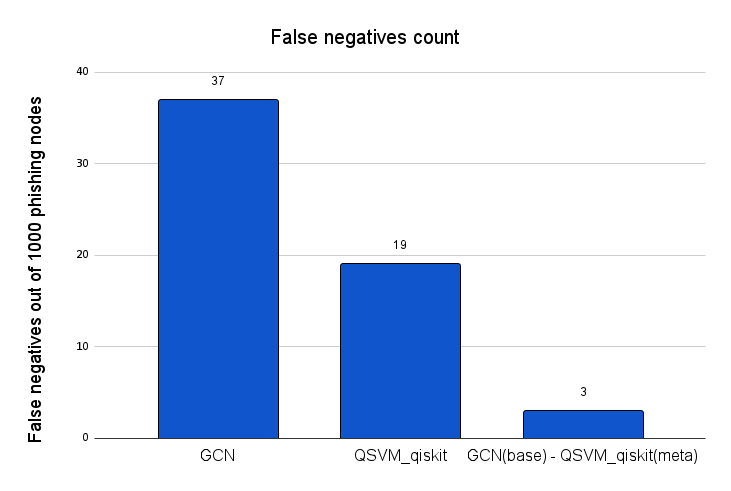}%
    \caption{False negatives count of individual models and best performing model combinations.}%
    \label{fig:fp}%
\end{figure}

The combination of two complementary algorithms QSVM\_qiskit (because of the least false negatives count) with GCN (because of overall best performance (Figure \ref{fig:indmodel}) but not so good at false negatives) generated the least false positives count than individual models as shown in Figure \ref{fig:fp}.

\subsection{QRAC results}

\begin{figure}[htb]
    \centering
    \includegraphics[width = 0.6\textwidth]{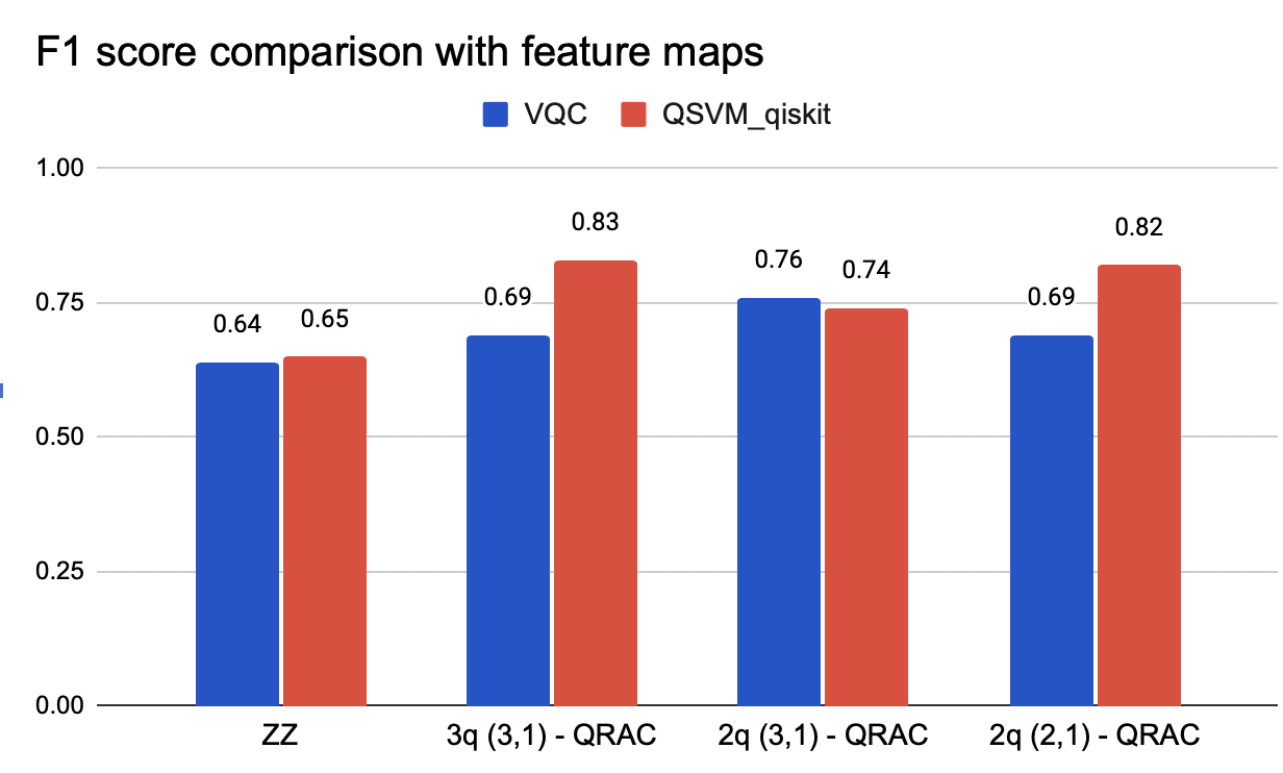}%
    \caption{Comparison of 2-qubit (2,1)-QRAC, 2-qubit (3,1)-QRAC and 3-qubit (3,1)-QRAC with ZZ feature map; Train data : 160p and 160np; Test data : 1000p and 10000np.}%
    \label{real vs binary}%
\end{figure}

Figure~\ref{real vs binary} represents the comparison of 2 qubit (3,1)-QRAC, 2 qubit (2,1)-QRAC and 3 qubit (3,1)-QRAC with ZZ feature map for VQC and QSVM\_qiskit. ZZ feature map used real valued features whereas QRACs have binarized data. We observed that 2 qubit (3,1)-QRAC performed better than other embeddings for VQC and 3 qubit (3,1)-QRAC is performing better for QSVM\_qiskit. One key finding is that the 2 qubit (2,1)-QRAC results closely matches that of 3 qubit (3,1)-QRAC. The both 2 qubit versions either produced enhanced results or similar results to that of 3 qubit version. This experiment emphasizes on a key argument that data compression can be efficiently done without losing essential information.

For hardware runs, we chose two feature encodings, that produced the best F1 score for VQC and QSVM. Although 2 qubit (2,1)-QRAC closely matched performance of 3 qubit (3,1)-QRAC for QSVM, the 3 qubit version had a slight edge.

\subsubsection{QRAC Hardware results}

\begin{table}[h]
    \caption{ VQC and QSVM results with QRAC on Quantum hardware and simulators on same train and test data}
    \centering
    \begin{tabular}{|c|c|c|c|c|c|c|c|c|}
    \hline
    \textbf{Models} & \textbf{Encoding} & \textbf{Device}  & \textbf{Macro F1} & \textbf{Phish\_F1} & \textbf{Phish\_Recall}\\
    \hline
    VQC & QRAC(3 qubits) & simulator & 0.78 & 0.82 & 0.96 \\
    VQC & QRAC(3 qubits) & ibm\_kingston & 0.75 & 0.80 & 0.95\\
    VQC & QRAC(2 qubits) & simulator & 0.89 & 0.90 & 0.96\\
    VQC & QRAC(2 qubits) & ibm\_kingston & 0.61 & 0.73 & 0.95\\
    QSVM\_qiskit & QRAC(3 qubits) & simulator & 0.94 & 0.94 & 0.95\\
    QSVM\_qiskit & QRAC(3 qubits) & ibm\_kingston & 0.91 & 0.91 & 0.90\\
    QSVM\_qiskit & QRAC(2 qubits) & simulator & 0.92 & 0.92 & 0.96\\
    QSVM\_qiskit & QRAC(2 qubits) & ibm\_kingston & 0.90 & 0.90 & 0.94\\
    \hline
    \end{tabular}
    \label{tab:VQChardware}
\end{table}
In Table \ref{tab:VQChardware}, all algorithms are trained on 320 data points and evaluated on 300 test data points. For hardware experiments, due to execution cost, we used a balanced subset of 300 test data points instead of 11,000 as used for simulator experiments and all experiments were executed using 128 shots for cost effectiveness. All circuits were transpiled using optimization level 1 and
no additional error mitigation was used. The results compare VQC and QSVM\_qiskit models using QRAC-based feature embeddings on both simulators and IBM quantum hardware.

From Table \ref{tab:VQChardware}, we observe that the QSVM\_qiskit models achieve the best overall performance across both simulator and hardware executions. In particular, QSVM\_qiskit with QRAC encoded into 3 qubits achieves a Macro F1 score of 0.94 on the simulator and maintains a high Macro F1 score of 0.91 on the ibm\_kingston hardware backend. Similarly, the 2-qubit QRAC-QSVM configuration achieves Macro F1 scores of 0.92 on the simulator and 0.90 on hardware, indicating strong robustness against hardware noise.

For the VQC models, the 2-qubit QRAC configuration performs best on the simulator with a Macro F1 score of 0.89 and phishing recall of 0.96. However, when executed on ibm\_kingston, the Macro F1 score drops significantly to 0.61, despite maintaining a high phishing recall of 0.95. A similar trend is observed for the 3-qubit QRAC-VQC setup, where the Macro F1 decreases from 0.78 on the simulator to 0.75 on hardware while preserving a phishing recall of 0.95. These results indicate that the VQC models remain highly sensitive to phishing samples even under hardware noise, resulting in low false negatives, which is an important property for anomaly and phishing detection tasks.

Another key observation is that the QSVM\_qiskit models exhibit much smaller degradation between simulator and hardware performance compared to VQC models. This suggests that kernel-based quantum models are currently more resilient to noise on near-term quantum hardware. The relatively stable hardware performance of QRAC-QSVM on ibm\_kingston demonstrates the effectiveness of QRAC embeddings for practical quantum kernel estimation on real devices, while the larger degradation observed in QRAC-VQC highlights the need for improved circuit optimization and error mitigation techniques for variational quantum models.

\section{Conclusion}

This study investigated the potential of QML algorithms, specifically VQC and QSVM, for phishing detection in Ethereum transaction networks. Through systematic benchmarking, we demonstrated the effectiveness of QRAC-based feature encoding, which consistently outperformed the conventional ZZ feature map. QRAC encoding improved VQC performance by up to 13\% and QSVM performance by up to 3\%, highlighting the benefits of efficient feature compression and qubit utilization. Furthermore, the proposed 2-qubit cascaded QRAC achieved performance comparable to, and in some cases exceeding, that of the 3-qubit QRAC. These results indicate that effective feature representations can be obtained using fewer qubits.

The proposed hybrid quantum-classical ensemble framework demonstrated the complementary strengths of QSVM and GCN. While GCN provided strong overall classification performance, QSVM contributed to reducing false negatives. By combining these models, the ensemble consistently improved phishing recall, a key metric for anomaly detection tasks where missed detections can have significant consequences.

Hardware experiments further demonstrated the feasibility of deploying QML models on current quantum processors. QSVM maintained performance close to simulator results for both 2-qubit and 3-qubit QRAC encodings, indicating strong robustness to hardware noise. VQC also showed good agreement between simulator and hardware results for the 3-qubit QRAC, while consistently maintaining high phishing recall across all hardware experiments. These findings highlight the potential of QRAC-based QML approaches for practical cybersecurity applications within the constraints of current NISQ devices.

Future work will focus on evaluating larger datasets, investigating advanced error mitigation strategies, and exploring homogeneous quantum ensemble architectures. Further theoretical analysis of cascaded QRAC encodings, together with their extension to higher-dimensional quantum systems, may provide deeper insights into resource-efficient QML for anomaly detection and related cybersecurity applications.

\backmatter

\section*{Declarations}
\subsection{Funding}
S.S.G. and A.P. acknowledge support from the Mphasis F1 Foundation, an Institute of Excellence grant funded by the Ministry of Education.
\subsection{Conflict of interest/Competing interests}
Not applicable
\subsection{Ethics approval} 
Not applicable
\subsection{Consent to participate}
Not applicable
\subsection{Consent for publication}
Not applicable
\subsection{Availability of data and materials}
The data that support the findings of this study are available from the authors upon request.
\subsection{Code availability}
Ensemble framework codes are publicly available in github \cite{gitcode}
\subsection{Authors' contributions} 
Conceptualization: Sai Sakunthala Guddanti, Anupama Ray; Methodology: Sai Sakunthala Guddanti, Anupama Ray, Mrunal Arun Kumavat; Formal analysis and investigation: Sai Sakunthala Guddanti, Anupama Ray, Anil Prabhakar, Mrunal Arun Kumavat; Writing - original draft preparation: Sai Sakunthala Guddanti; Writing - review and editing: Anupama Ray, Anil Prabhakar; Funding acquisition: Anil Prabhakar; Resources: Anil Prabhakar; Supervision: Anupama Ray, Anil Prabhakar.
\begin{appendices}




\end{appendices}


\bibliography{refs}

\end{document}